**Magnetic properties of layered hybrid organic-inorganic metal-halide perovskites: transition metal, organic cation and perovskite phase effects**

*Yaiza Asensio, Sergio Marras, Davide Spirito, Marco Gobbi, Mihail Ipatov, Fèlix Casanova, Aurelio Mateo-Alonso, Luis E. Hueso\* and Beatriz Martín-García\**

Y. Asensio, M. Gobbi, F. Casanova, L.E. Hueso, B. Martín-García
CIC nanoGUNE BRTA, Tolosa Hiribidea, 76, 20018 Donostia-San Sebastián, Basque Country, Spain
Email: l.hueso@nanogune.eu, b.martingarcia@nanogune.eu

S. Marras
Istituto Italiano di Tecnologia - Materials Characterization Facility, Genova 16163, Italy

D. Spirito
IHP–Leibniz-Institut für innovative Mikroelektronik, Im Technologiepark 25, 15236 Frankfurt (Oder), Germany

M. Gobbi
Materials Physics Center CSIC-UPV/EHU, 20018 Donostia-San Sebastián, Spain

M. Ipatov
SGIker Medidas Magnéticas Gipuzkoa, UPV/EHU, 20018 Donostia-San Sebastián, Spain

A. Mateo-Alonso
POLYMAT, University of the Basque Country UPV/EHU, Avenida de Tolosa 72, 20018 Donostia-San Sebastián, Spain

M. Gobbi, F. Casanova, A. Mateo-Alonso, L.E. Hueso
IKERBASQUE, Basque Foundation for Science, 48009 Bilbao, Spain

ABSTRACT. Understanding the structural and magnetic properties in layered hybrid organic-inorganic metal halide perovskites (HOIPs) is key for their design and integration in spin-electronic devices. Here, we have conducted a systematic study on ten compounds to understand the effect of the transition metal ($Cu^{2+}$, $Mn^{2+}$, $Co^{2+}$), organic spacer (alkyl- and aryl-ammonium) and perovskite phase (Ruddlesden-Popper and Dion-Jacobson) on the properties of these materials. Temperature-dependent Raman measurements show that the crystals' structural phase transitions are triggered by the motional freedom of the organic cations as well as by the flexibility of the inorganic metal-halide lattice. In the case of $Cu^{2+}$ HOIPs, an increase of the in-plane anisotropy and a reduction of the octahedra interlayer distance is found to change the behavior of the HOIP from that of a 2D ferromagnet to that of a quasi-3D antiferromagnet. $Mn^{2+}$ HOIPs show inherent antiferromagnetic octahedra intralayer interactions and a phenomenologically rich magnetism, presenting spin-canting, spin-flop transitions and metamagnetism controlled by the crystal anisotropy. $Co^{2+}$ crystals with non-linked tetrahedra show a dominant paramagnetic behavior irrespective of the organic spacer and the perovskite phase. This work demonstrates that the chemical flexibility of HOIPs can be exploited to develop novel layered magnetic materials with tailored magnetic properties.

## 1. Introduction

A growing number of magnetic layered (2D) materials are being studied since the recent isolation of intrinsically magnetic monolayers. Today the list includes transition metal chalcogenides (*e.g.* $Cr_2Ge_2Te_6$ and $Fe_3GeTe_2$), metal halides (*e.g.* $CrI_3$), and metal phosphorus trichalcogenides (MPS$_3$, M = $Fe^{2+}$, $Co^{2+}$, $Ni^{2+}$, $Mn^{2+}$)[1–3], as well as less explored covalent organic frameworks[4] or inorganic-organic hybrid materials, including metal organic frameworks,[5–9] metal-organic crystals,[10] organically intercalated layered materials,[11,12] and hybrid organic-inorganic metal-halide perovskites (HOIPs)[13,14]. One of the goals of the research around 2D magnets is to achieve control over their magnetic properties. [3,15–17] In this regard, layered transition metal HOIPs offer an ideal platform for the engineering of magnetic properties thanks to their chemical and structural versatility. These HOIPs consist of anionic inorganic metal-halide sheets separated by organic ammonium-based cations. The metal-halide (MX) sheets can take on very different compositions, with M = $Cu^{2+}$, $Cr^{2+}$, $Mn^{2+}$, $Fe^{2+}$, $Co^{2+}$ and X = Cl, Br. As for the HOIP's crystal structure, depending on whether the organic cations are monovalent or divalent, the HOIP will exhibit a Ruddlesden-Popper (RP) or Dion-Jacobson (DJ) perovskite phase, respectively.[14,18] Lastly, like graphene and other similar 2D materials, layered HOIPs can be mechanically exfoliated.[19,20]

Layered HOIPs, especially Pb-based compounds, have been extensively studied due to their outstanding optoelectronic properties[21–23]. So far, comparatively few works have sought to explore and understand their magnetic properties[14]. Those that have, have revealed a rich phenomenology: depending on the transition metal ion ($Cu^{2+}$ [24–36], $Cr^{2+}$ [13], $Mn^{2+}$ [28,37–41], $Fe^{2+}$ [42–45], Co$^{2+}$ [28,46], Ru$^{3+}$[47] and Mo$^{3+}$[47]), the HOIP can exhibit ferromagnetic (FM), antiferromagnetic (AFM), coexistence of FM/AFM,[33] paramagnetic or no magnetic ordering at all[47]. In contrast, other hybrid perovskite-type magnetic materials such as metal-formates[14,48–50] and hypophosphites[51] have shown less magnetic tunability by chemical design. They show spin-canted antiferromagnetism regardless of the transition metal, the organic cation and the anion.[14,48–51]

Research efforts on hybrid metal-halide perovskites have mainly focused on RP perovskite phases, yet RP and DJ phases are drastically different in terms of crystal packing direction and interlayer distance. These differences translate into local lattice distortions and variations in the length and angles of the X···M···X bonds, which in turn affect the resulting magnetic behavior.[14,18,45,52] Moreover, the hydrogen bonding between the inorganic and organic parts in layered HOIPs induces a complex interaction between structural deformation of the octahedra and the conformational flexibility of the organic cations, leading to the appearance of structural phase transitions.[53–56] In some cases, the emerging crystal phases exhibit new properties, such as ferroelectricity[26,57] and ferroelasticity[58]. Therefore, it is important to determine the presence/absence of the structural phase transitions, considering their relevance in terms of materials' properties and thermal stability for a future integration of these materials in devices.

Because of their variety, these materials are promising candidates for on-demand magnetism by chemical design. Indeed, a library of layered HOIPs and their magnetic properties would provide an invaluable tool to explore their use in optoelectronic and spintronic devices, but this is not yet available. In this work, we take a step in this direction and provide a systematic and comprehensive study of the influence of the metal, the organic spacer and the perovskite phase on the magnetic properties of layered transition metal HOIPs. We use temperature-dependent micro-Raman spectroscopy to determine the materials' structural phase transitions, and, importantly, we show that the number and nature of the detected phase transitions varies with the composition of the organic moiety and the metal halide. We also report on the temperature-dependent photoluminescence properties of the Mn$^{2+}$ compounds under study. Furthermore, we report for the first time on the magnetic behavior of five compounds: p-PEAACuCl$_4$ (($NH_3C_6H_4NH_3$)$^{2+}$ referred to as PEAA), EA$_2$MnCl$_4$, (($C_2H_5NH_3$)$^+$ referred to as EA), EDAMnCl$_4$ (($NH_3C_2H_2NH_3$)$^{2+}$ referred to as EDA), PEA$_2$CoCl$_4$ (($C_6H_5CH_2CH_2NH_3$)$^+$ referred to as PEA) and EDACoCl$_4$. We show that Cu$^{2+}$, Mn$^{2+}$ and Co$^{2+}$ crystals have very different magnetic behavior. In the case of Cu$^{2+}$ HOIPs, modulating the in-plane anisotropy and the interlayer distance changes the HOIP's behavior from that of a 2D ferromagnet to that of a 3D antiferromagnet. In contrast, for the Mn$^{2+}$ HOIPs, it is the anisotropy of the octahedral lattice, caused by the organic cation conformational rearrangement associated with the perovskite phase and the moiety-halide hydrogen bonding, which impacts the magnetic properties, keeping an AFM behavior and leading to the appearance of spin-canting, spin-flop or even metamagnetic phenomena. In $Co^{2+}$ hybrid materials, with a non-perovskite structure, a paramagnetic behaviour is observed, while the magnitude of the magnetization can be modulated by properly choosing the perovskite phase and the interlayer distance.

## 2. Results and Discussion
### 2.1. Materials Characterization

We synthesized layered crystals using as inorganic precursors metal (II) chlorides ($CuCl_2$, $MnCl_2$ and $CoCl_2$) mixed with the corresponding mono- or di-ammonium cations in polar solvents or an acid medium to obtain the Ruddlesden-Popper (RP) or Dion-Jacobson (DJ) perovskite phases, respectively (see procedures in the Experimental section). To evaluate the effect of the organic moieties, we selected alkyl and phenyl cations. Specifically, we used: phenylethylammonium (PEA) and ethylammonium (EA) for the RP crystals and p-phenylenediammonium (PEAA) and ethylenediammonium (EDA) for the DJ compounds. While the alkylammonium chains provide flexibility due to their higher conformational and motional freedom, the phenyl rings offer more rigidity due to π–π stacking in the organic layer and their larger volume. Therefore, each of these molecules will cause a different distortion of the inorganic lattice.[18] We obtained $(R-NH_3^+)_2MCl_4$ RP and $(NH_3^+-R-NH_3^+)MCl_4$ DJ bulk crystals, which consist of single octahedra ($CuCl_6^{4-}$, $MnCl_6^{4-}$)/tetrahedra ($CoCl_4^{2-}$) layers separated by two or one layer of organic cations, respectively (see **Figure 1** and Figures S1-2 for the other materials). Since the $Co^{2+}$ compounds do not show a metal-halide octahedral coordination, they cannot be considered perovskites.[59] Therefore, hereafter we will refer to them as HOI materials (HOIMs). We planned to synthesize and characterize four types of compounds: $PEA_2MCl_4$, $EA_2MCl_4$, $PEAAMCl_4$ and $EDAMCl_4$, with M= $Cu^{2+}$, $Mn^{2+}$ and $Co^{2+}$. All but two of the twelve compounds were successfully synthesized, the exceptions being $PEAAMnCl_4$ and $PEAACoCl_4$. These crystals did not crystallize due to the rigidity of the $MnCl_6^{4-}$ octahedra and $CoCl_4^{2-}$ tetrahedra framework in comparison with the distorted and flexible Jahn-Teller structure of $CuCl_6^{4-}$ octahedra sheets.[18,27,52] The four $Cu^{2+}$ crystals are shown in Figure 1a. X-ray diffraction (XRD) patterns of the different crystals confirmed the formation of the target compounds and their layered structure (see Figure 1c and Figures S1c-S2c for more data). From the periodicity of the diffraction peaks, we estimated the distance between the inorganic layers (d), which decreases from PEA >> EA > PEAA > EDA (*e.g.* in $Cu^{2+}$ from 19 Å → 8 Å, see Figure 1b) and also when going from an octahedral ($Cu^{2+}$, $Mn^{2+}$) to a tetrahedral ($Co^{2+}$) lattice (Table S1).

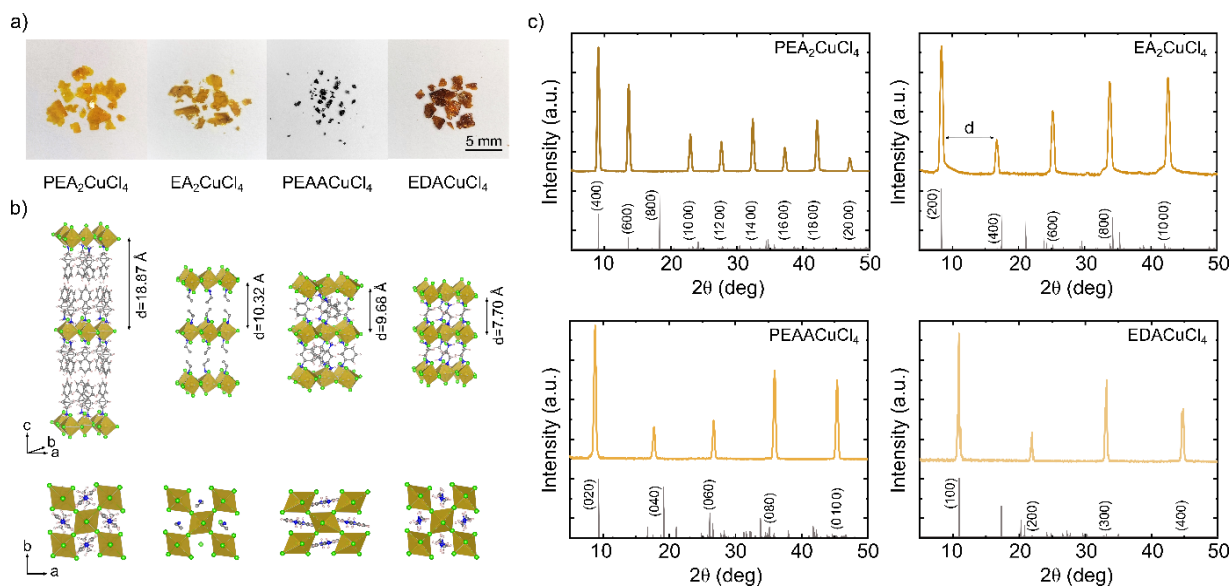

**Figure 1.** (a) Photographs of the Cu-based layered HOIPs with (b) schemes of their crystal structure drawn using VESTA software[60] using the available crystallographic data: $PEA_2CuCl_4$,[57] $EA_2CuCl_4$,[61] m-$PEAACuCl_4$,[32] and $EDACuCl_4$[62]. The interlayer distances *d* are obtained from the experimental XRD patterns shown in (c). The reference patterns for each compound are also presented as gray bars: $PEA_2CuCl_4$ (CCDC 751958), $EA_2CuCl_4$ (CCDC 1147994), m-$PEAACuCl_4$ (CCDC 2057987), and $EDACuCl_4$ (CCDC 1148696). XRD patterns displayed correspond to measurements from a representative crystal for each material.

Additionally, we determined the structural phase transitions of the layered crystals using temperature-dependent Raman spectroscopy. We monitored the vibrational modes of the inorganic lattice and the organic cations in terms of peak position and linewidth in the range 80-340 K[55,56,63–65]. We observed structural phase transitions only in the samples containing $EA^+$ (**Figure 2**a-c for samples with $EA^+$, and Figures S3-S20 for the rest of samples and detailed data analysis). In the case of the $Cu^{2+}$ HOIPs, we detected structural phase transitions for the $EA_2CuCl_4$ crystal at 230-250K and ~330K (Figure 2a), evidenced by changes in the 180 $cm^{-1}$ (in-plane Cu-Cl bending[66,67]), 250 $cm^{-1}$ (octahedra symmetric Cu-Cl stretching[66,67]) and 980 $cm^{-1}$ (C-N stretching[68,69]) modes due to rearrangements of the octahedra and $EA^+$ molecules. In accordance with XRD measurements,[70] the structural phase transitions follow the sequence from triclinic → orthorhombic → monoclinic[26]. For the $Mn^{2+}$ HOIPs, we observed one structural phase transition for the $EA_2MnCl_4$ crystal at ~230K only in the organic cation vibrations (290 $cm^{-1}$ $NH_3^+$ torsion, 872/978 $cm^{-1}$ C-C/C-N stretching, 1460/1500 $cm^{-1}$ $CH_3$ deformation and CH$_2$ scissoring and 1570 cm$^{-1}$ NH$_3^+$ related modes)[68,69,71] while the symmetric Mn-Cl stretching at the octahedra (210 cm$^{-1}$) showed no marked changes (Figure 2b). This is in accordance with the orthorhombic to orthorhombic structural transition *Pbca → Cmca (Abma)* reported by single crystal XRD measurements[72], which keeps the same crystal system and is attributed to a change in the molecules' arrangement.

For the Co$^{2+}$ HOIMs, the EA$_2$CoCl$_4$ crystal showed a structural phase transition at ~240K, evidenced by slight changes in the C-C stretching of the EA$^+$ molecules[71] (~866 cm$^{-1}$) accompanied by shifts in the symmetric Co-Cl stretching in the tetrahedra[73] (~275 cm$^{-1}$) (Figure 2c). This transition is ascribed to the orthorhombic to orthorhombic structural transition *Pnma → P2$_1$2$_1$2$_1$* based on reported single crystal XRD data[46]. Therefore, in this case the rearrangement of the molecules during the structural phase transition affects tetrahedral distortion.[55] By comparing the different crystals, it is noteworthy that the Jahn-Teller distortion of the Cu$^{2+}$ octahedral lattice leads to structurally more flexible systems that can accommodate a large number of differently sized organic cations and with dynamic structural phase transitions involving both the inorganic lattice and organic moieties, while this effect is less marked for the Co$^{2+}$ tetrahedral coordination. In contrast, the Jahn-Teller effect is absent in the Mn$^{2+}$ octahedral framework, resulting in lower structural flexibility. From this we conclude that in this case the structural phase transitions are associated with a rearrangement of the organic moieties.

Regarding the Mn$^{2+}$ HOIPs, these present the characteristic red photoluminescence (PL) emission associated with octahedrally coordinated Mn$^{2+}$ compounds resulting from the $^4T_{1g}$(G) → $^6A_{1g}$ (S) electronic transition between the triple-degenerate (T) and nondegenerate (A) levels (Figure 2d, see Figure S21 for PL spectra).[52,74,75]. The PL emission position and linewidth (full width half maximum - FWHM) are almost unaffected by the cation, with similar values for the three materials (PEA$^+$ ~ 607 nm / FWHM ~ 57 nm; EA$^+$ ~ 608 nm / FWHM ~ 61 nm; EDA$^+$ ~ 607 nm / FWHM ~ 55 nm at 300K). This indicates that the organic cations hardly modify the Mn-Cl orbitals or therefore, the band structure. The observed PL emission values compare well with those in the literature, *e.g.* MA$_2$MnCl$_4$ (MA = methylammonium) with 597-608 nm and FWHM ~ 72 nm.[74] Additionally, the structural phase transitions are also reflected in the trend of the PL emission with the temperature. In the case of the EA$_2$MnCl$_4$ crystal we observe a slight change in the slope of PL peak position and linewidth with temperature at ~230K.

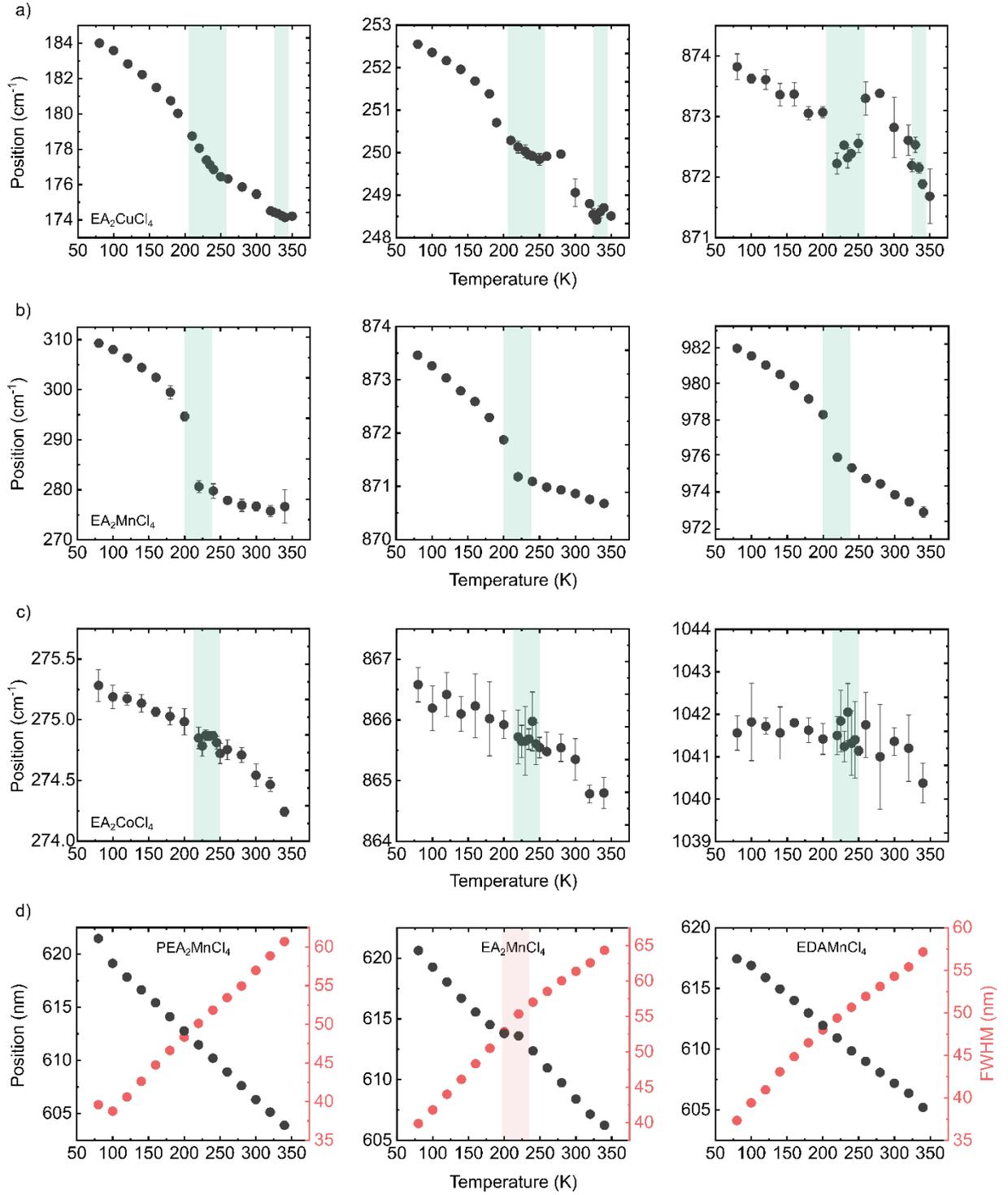

**Figure 2.** (a-c) Temperature dependence (80-350K range) of the peak position of selected Raman modes. (a) EA$_2$CuCl$_4$: in-plane Cu-Cl bending (~180 cm$^{-1}$),[66,67] symmetric Cu-Cl octahedra stretching (~250 cm$^{-1}$) [66,67] and C-C stretching (~872 cm$^{-1}$)[71]; (b) EA$_2$MnCl$_4$ : NH$_3^+$ torsion (~290 cm$^{-1}$),[71] C-C stretching (~872 cm$^{-1}$),[71] and C-N stretching (~978 cm$^{-1}$),[68,69]; and (c) EA$_2$CoCl$_4$: symmetric Co-Cl stretching (~275 cm$^{-1}$)[73], C-C stretching (~866 cm$^{-1}$)[71] and C-N/C-C stretching (~1042 cm$^{-1}$)[68,69]. (d) Temperature-dependent photoluminescence emission in terms of peak position and linewidth (full width half maximum - FWHM) for the $Mn^{2+}$ HOIP crystals. The regions shaded in green (Raman) and red (PL) indicate the temperature at which the structural phase transitions take place. Results displayed correspond to the average and standard deviation (error bars) determined from 5 points measured in a representative crystal for each material.

## 2.2. Magnetic Behavior

Composition and perovskite phase should play a role not only in the structural phase transitions but also in the resulting magnetic properties. The key element for the appearance of magnetic properties in these layered HOIMs is the transition metal incorporated in the crystal structure. Therefore, in the discussion below we compare systems with the same metal and different organic cation and perovskite phase.

In the case of the $Cu^{2+}$ HOIPs, we present the temperature-dependent magnetization curves, M(T), for the $PEA_2CuCl_4$ RP crystal (**Figure 3a**). The Curie-Weiss fittings (see Figure S22 for representative examples) yield positive temperature intercept values (θ) for in-plane and out-of-plane measurements, indicating predominant ferromagnetic (FM) Cu-Cu spin intra- and inter-layer interactions. This is also supported by the positive values of the intra- and inter-layer exchange constants ($J/k_B$, $J'/k_B$) determined using the quadratic layer series expansion of Baker *et al.*[25,76] (see SI for further fitting details and Table S3 for magnetic parameters). Most importantly, long-range FM order is achieved at $T_C$ ~ 12 K, which matches well with reported values for this crystal[24,30,38,39]. The in-plane and out-of-plane field-dependent magnetization curves, M(H), saturate at ~$1\mu_B$, although with different saturation fields $H_S$ ~100 Oe and $H_S$ ~1500 Oe, respectively (Figure 3e). These measurements highlight the FM behavior of this crystal with an easy axis parallel to the $[CuCl_4]^{2-}$ layer. Indeed, the intralayer FM nature in these $Cu^{2+}$ HOIPs has been attributed to the existence of nearly orthogonal (non-overlapping) magnetic *d* orbitals on corner-sharing $[CuCl_4]^{2-}$ octahedra resulting from their Jahn-Teller distortion.[52] This can be structurally related to the presence of Cu···X···Cu angles > 160 deg in the inorganic lattice, as occurs in the $PEA_2CuCl_4$ with ~180 deg.[27,52] Regarding the interlayer interactions, which are driven by dipolar interactions through the apical Cl- in these $Cu^{2+}$ HOIPs,[27,52] these should be weaker due to the large interlayer distance, ~20 Å, leading to $J'/J < 10^{-4}$ and therefore to a 2D ferromagnetic behavior ($J'/J < 10^{-3}$)[27,52]. In the case of the $EA_2CuCl_4$ RP crystal, the M(T) curves show a $T_C$ ~ 11K [25,26] (Figure 3b). While the Curie-Weiss fittings show θ > 0 for in-plane and out-of-plane measurements, indicating FM interactions in both directions, the $J/k_B > 0$ and $J'/k_B < 0$ show that there are FM interactions in the $[CuCl_4]^{2-}$ plane while antiferromagnetic (AFM) interactions appear in the perpendicular direction (Table S3). Therefore, the small kink observed at around 10K can be ascribed to weak AFM out-of-plane interactions.[25,26] Furthermore, the M(H) curves in both crystal directions are similar to those of the $PEA_2CuCl_4$ crystal (Figure 3f). In this case, the saturation fields ($H_S$) are larger than for $PEA_2CuCl_4$, *i.e.*, for in-plane $H_S$ ~1300 Oe and for out-of-plane $H_S$~2700 Oe. Indeed, M(T) curves acquired at higher field show a clear FM behavior in both directions (Figure S23), although the magnetization reached is still smaller than for the $PEA_2CuCl_4$ crystal. This magnetic behavior can be explained by looking into the crystal structure. In the $EA_2CuCl_4$ HOIP the in-plane Cu···X···Cu angle has a value of ~170 deg (smaller than in $PEA_2CuCl_4$) and the interlayer distance is significantly reduced to ~11 Å (almost half of that of $PEA_2CuCl_4$); J'/$k_B$ is negative and one order of magnitude higher than in the $PEA_2CuCl_4$ crystal. This creates a scenario where the FM interactions in the $[CuCl_4]^{2-}$ octahedra plane still dominate but with AFM contributions from the interlayer interaction, keeping a 2D ferromagnetic behavior overall.

In the case of the $PEAACuCl_4$ DJ HOIP, the changes in the magnetic behavior begin to be more marked. The in- and out-of-plane M(T) curves display a sharp increase up to a broad maximum at $T_C$ ~ 13K, with magnetizations one order of magnitude lower than in the previously shown $Cu^{2+}$ RP HOIPs (Figure 3c). Although the Curie-Weiss fittings show θ > 0 in both directions, the J/$k_B$ > 0 and J'/$k_B$ < 0 and most importantly a J'/$k_B$ ~ -$10^{-2}$ (ten-times higher than in $EA_2CuCl_4$ HOIP) show that ferromagnetism is preserved in the $[CuCl_4]^{2-}$ plane with stronger AFM interactions appearing in the perpendicular direction (Table S3). Therefore, the intralayer FM interactions compete with the interlayer AFM interactions to govern the magnetic behavior. Furthermore, the M(H) curves tend to saturate and achieve the FM state at larger fields compared to RP crystals (Figure 3g). Indeed, M(T) curves acquired at 10 kOe show FM behavior in both directions (Figure S23), with magnetizations clearly smaller than for the RP crystals under study. The combination of the DJ perovskite phase and a bulky aromatic cation in this HOIP leads to higher anisotropy in-plane with smaller Cu···X···Cu angles (~160 deg) and a shorter interlayer distance (~9.5 Å), with the J'/$k_B$ two or one order of magnitude larger compared to the $PEA_2CuCl_4$ and $EA_2CuCl_4$ crystals, respectively. Therefore, the FM interactions in the $[CuCl_4]^{2-}$octahedra plane still stand out, maintaining a 2D ferromagnetic behavior (J'/J ~ $10^{-3}$)[27,52] but with AFM contributions from the interlayer interaction becoming more relevant.

The $EDACuCl_4$ DJ crystal is where the magnetic changes related to its structure are more clearly reflected. The in- and out-of-plane M(T) curves display a slow increase in the magnetization, reaching a broad maximum at $T_C$ ~ 36K, with values two orders of magnitude lower than in the PEAACuCl$_4$ DJ HOIP (Figure 3d). It is worth noting that θ > 0 and J/k$_B$ > 0 indicate the presence of FM intralayer interactions, while J'/k$_B$ ~ -1 (two orders of magnitude higher than PEAACuCl$_4$ crystal) shows that the AFM interactions occurring in the perpendicular direction are significantly stronger. Therefore, the interlayer AFM interactions govern the magnetic behavior in this crystal. Moreover, the M(H) curves show a linear increase with the field without reaching saturation, as observed in weak ferromagnets (Figure 3h). Furthermore, M(T) curves acquired at 10 kOe demonstrate that the AFM behavior is also kept in both directions. even at this field (Figure S23), and therefore, the FM state is not reached. The short interlayer distance (~8 Å) combined with a higher in-plane anisotropy from Cu···X···Cu angles (~160 deg) and more pronounced octahedral tilting favor the AFM contributions from the interlayer interaction, resulting in J'/J ~ $10^{-1}$, closer to a 3D antiferromagnet rather than a 2D ferromagnet (J'/J < $10^{-3}$)[27,52].

Comparing all Cu$^{2+}$ HOIPs, there is a systematic interlayer reduction PEA$^+$ > EA$^+$ > PEAA$^+$ > EDA$^+$ (20 Å → 8 Å) accompanied by a closing in the Cu···Cl···Cu angles on corner-sharing [CuCl$_4$]$^{2-}$octahedra from 180 deg (PEA$^+$) to 167 deg (EDA$^+$) (Tables S1-S2). This structural change leads to AFM interlayer contributions (driven by the dipolar interactions through the apical Cl$^-$[27,52]) being greater than the intralayer FM interactions (originating from non-overlapping orthogonal magnetic *d* orbitals favored by Cu···X···Cu angles close to 180 deg[52]), and therefore to a change in behavior from that of a 2D ferromagnet in PEA$_2$CuCl$_4$ (J'/J < $10^{-4}$) to a 3D antiferromagnet in EDACuCl$_4$ (J'/J ~ $10^{-1}$).

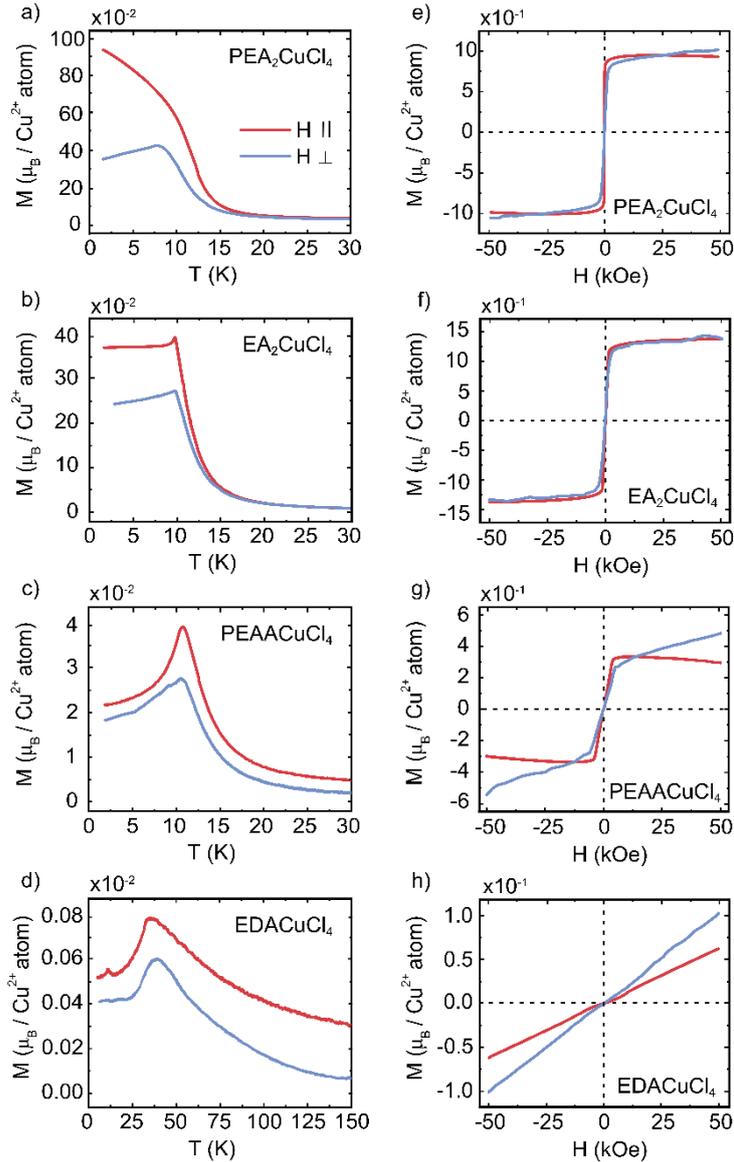

**Figure 3.** Magnetization (M) vs temperature (T) at 500 Oe parallel (∥, red lines) and perpendicular (⊥, blue lines) to the $[CuCl_6]^{4-}$ octahedra layers for (a) $PEA_2CuCl_4$, (b) $EA_2CuCl_4$, (c) $PEAACuCl_4$ and (d) $EDACuCl_4$. Magnetization (M) vs applied field (H) at 5 K parallel (∥, red lines) and perpendicular (⊥, blue lines) to the $[CuCl_6]^{4-}$ octahedra layers for (e) $PEA_2CuCl_4$, (f) $EA_2CuCl_4$, (g) $PEAACuCl_4$ and (h) $EDACuCl_4$. Results shown correspond to M(T) and M(H) curves collected from a representative crystal for each material.

In the case of the $Mn^{2+}$ HOIPs, the M(T) curves also clearly show different magnetic behavior depending on the organic cation and perovskite phase (**Figure 4**a-c and SI Figures S26-27 for angle-dependent measurements). The Curie-Weiss fittings yield large and negative θ for in-plane and out-of-plane measurements and $J/k_B < 0$ in all the structures (Table S4).[37,40] The intralayer AFM nature in the $Mn^{2+}$ HOIPs is ascribed to the overlap of the magnetic *d* orbitals on adjacent [MnCl$_4$]$^{2-}$octahedra through the orbital of the corner-shared Cl$^-$, favored by the almost linear and symmetrical Mn···Cl···Mn bonds.[52] However, differences arise among the different compositions. The PEA$_2$MnCl$_4$ RP crystal shows a broad maximum in the M(T) at ~87K, while a sharp change in slope with a steep increase of the magnetization takes place at the Neel temperature, T$_N$ ~47K,[37,39,40] being more marked in the in-plane direction (Figure 4a). Since the magnetization tends to drop toward zero in the out-of-plane direction, the easy axis for the AFM alignment is the perpendicular direction to the [MnCl$_6$]$^{4-}$ octahedra plane. The change in the slope in the in-plane M(T) curve is ascribed to the appearance of a weak FM component arising due to spin-canting occurring at the plane from antisymmetric Dzyaloshinski-Moriya interactions related to the tilting of the [MnCl$_6$]$^{4-}$ octahedra in the crystal structure (see Figure S27 for angle-dependence M(T)).[37,40] Additionally, the M(H) curve (Figure 4d) shows a magnetic hysteresis loop driven by the interplay between the easy axis and the antisymmetric Dzyaloshinsky–Moriya interactions ascribed to the distortion of the crystal structure.[37,40] In the out-of-plane M(H) curve at 5K, we observe a spin-flop transition at ~25 kOe, confirming that the easy axis for AFM ordering is perpendicular to the [MnCl$_6$]$^{4-}$ octahedra lattice[37,40]. In fact, the collected out-of-plane M(T) curves at high fields show characteristic AFM behavior.[37] From the M(H) curves, we determine the remanent magnetization (M$_{rem}$=5.6·10$^{-3}$ μ$_B$ /Mn$^{2+}$atom) from which we estimate the spin-canting angle using the expression M$_{rem}$=M$_s$sinφ[77]. The obtained value, φ=0.06 deg, matches well with reported values for the PEA$_2$MnCl$_4$ crystal[37,40].

In the case of the EA$_2$MnCl$_4$ RP crystal, the in-plane M(T) curve shows a similar behavior to that of the PEA$_2$MnCl$_4$ crystal (a broad maximum at ~80K and T$_N$~45K), while the out-of-plane curve does not (Figure 4b). These data indicate that a weak FM behavior in both directions appears due to spin-canting. In the M(H) curves we observe a more complex magnetic behavior in comparison to the PEA$_2$MnCl$_4$ crystal, especially in the direction perpendicular to the [MnCl$_6$]$^{4-}$ octahedra plane (Figure 4e, see also Figure S26-27 for angle-dependent measurements). The in-plane M(H) curve presents a weak ferromagnetic behavior, and from the remanent magnetization 2.5·10$^{-3}$ μ$_B$ /Mn$^{2+}$atom, we estimated a spin-canting angle value of 0.03 deg, slightly lower than for PEA$_2$MnCl$_4$. The out-of-plane M(H) curve shows different regimes. The sudden rise in the range 300-1500 Oe corresponds to a metamagnetic transition from an AFM state to a weak FM state due to the emergent spin-canting[38], while the inflection point at ~25 kOe fits with a spin-flop transition[37,40]. Considering these transitions and the out-of-plane M(T) at different magnetic fields (Figure S25), there are two contributions at different fields: at 100 Oe the crystal is in the same regime as PEA$_2$MnCl$_4$ with a predominant antiferromagnetism, while at 500 Oe (Figure 4b) there is a FM contribution which explains the jump in the magnetization at ~45K. The AFM state is recovered at higher field, as evidenced by the M(T) curve at 1T, while above 25 kOe the magnetization shows FM-like behavior as a consequence of the spin-flop transition (Figure S25). It is important to note that the coexistence of metamagnetism, spin-canting and spin-flop phenomena in the same crystal is not common, since a suitable combination of crystal structure, anisotropy and nearest/next-nearest interactions is required[38]. Indeed, it has not been reported in layered HOIPs before.

When passing from the $Mn^{2+}$ RP structures to the $EDAMnCl_4$ DJ crystal (Figure 4c), both the in-plane and out-of-plane M(T) curves show an AFM behavior with a broad maximum at $T_N$~80K, reaching a long-range ordering state. Similar to the other $Mn^{2+}$ HOIPs, the easy axis remains in the direction perpendicular to the $[MnCl_4]^{2-}$ octahedra plane. Importantly, there is no spin-canting in this DJ crystal. The M(H) curves show the same linear trend for the in-plane case, but no hysteresis, and a less marked spin-flop transition at ~25kOe in the out-of-plane case, *i.e.*, the easy axis (see Figure S24 for M(T) at high fields showing the FM-like behavior). Therefore, the absence of spin-canting and a subtle spin-flop transition indicate that this DJ crystal presents a smaller anisotropy than the RP structures, while a large anisotropy is present in the phenomenologically rich $EA_2MnCl_4$ crystal.

When comparing the $Mn^{2+}$ RP and DJ structures (Table S2), we observe that the Mn⋯Cl⋯Mn angles and distances allow the overlap of the magnetic *d* orbitals, giving rise to the intralayer antiferromagnetism[52] observed in all of them. However, looking into the crystal structures, we see that the interlayer octahedra arrangement is different in the RP and the DJ compounds, producing a different symmetry among the metal atoms in-plane and especially out-of-plane. As commented above, the source of the spin-canting in $Mn^{2+}$ compounds is the antisymmetric Dzyaloshinski-Moriya interactions that are related not only to the tilting of the bounded $[MnCl_6]^{4-}$ octahedra in the plane,[37,40] but also to the structural anisotropy between the $Mn^{2+}$ ions present in the crystal structure.[78,79] The latter can be the explanation for the absence of spin-canting in the DJ crystal. As can be seen in Figure S1b, in the RP crystals $[MnCl_6]^{4-}$ octahedra in neighboring planes are staggered relative to one another, while in the DJ they are perfectly aligned. The presence of symmetry between the $Mn^{2+}$ ions together with a shorter interlayer distance in the DJ $EDAMnCl_4$ crystal should alter the spin structure leading to the absence of spin-canting.

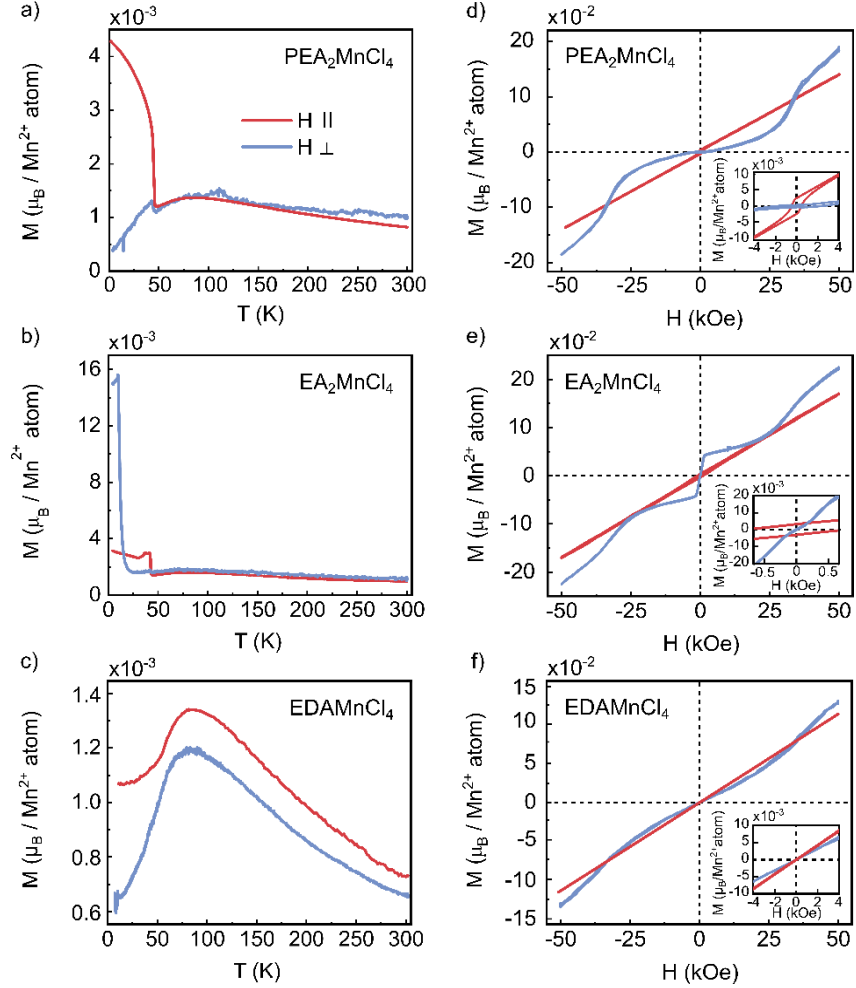

**Figure 4.** Temperature (T) dependent magnetization (M) at 500 Oe parallel (∥, red lines) and perpendicular (⊥, blue lines) to the $[MnCl_6]^{4-}$ octahedra layers for (a) $PEA_2MnCl_4$, (b) $EA_2MnCl_4$ and (c) $EDAMnCl_4$. Field (H) dependent magnetization (M) at 5 K parallel and perpendicular to the $[MnCl_6]^{4-}$ octahedra layers for (d) $PEA_2MnCl_4$, (e) $EA_2MnCl_4$ and (f) $EDAMnCl_4$. The insets show the zoom-in of the M(H) curves at low magnetic fields. Results shown correspond to M(T) and M(H) curves collected from a representative crystal for each material.

In the case of the $Co^{2+}$ HOIMs, the M(T) curves show a paramagnetic behavior regardless of the organic cation and the perovskite phase (**Figure 5**a-c). This behavior can be ascribed to the isolated nature of the $[CoCl_4]^{2-}$ tetrahedra (no shared corner) in the crystal structure (Figure S2)[46]. The Curie-Weiss fittings yield a small negative θ for in-plane measurements in the RP-like compounds, indicating predominant weak AFM interactions between the spins at the intralayer Co sites, while a small positive value is observed for out-of-plane measurements, pointing to weak interlayer Co-Co spin FM interactions (Table S5). The intralayer weak AFM Co-Co spin interactions are also confirmed by the negative and small value of $J/k_B$. In contrast, for the DJ-like crystal, θ is positive and small in both cases, in-plane and out-of-plane, suggesting weak Co-Co spin FM interactions, although the fact that $J/k_B < 0$ indicates the existence of AFM interactions as well, which could explain the small kink observed in the M(T) curves. The effective magnetic moment, $\mu_{eff}$, obtained from the same fit is larger than the spin-only moment, $\mu_{SO} = 3.78\ \mu_B$, in all cases, confirming the presence of orbital contributions.[46] Interestingly, the M(H) data collected at 5K for the RP-like crystals (PEA and EA) show that there is spin flipping at high fields[46] both in the in-plane and out-of-plane cases. This is not observed for the EDA-DJ crystal, indicating that the paramagnetic contribution is more marked in the latter (Figure 5d-f). Moreover, the maximum moment reached at 4.5T for the in-plane and out-of-plane measurements follows the trend PEA > EDA > EA, which matches well with the Co-Co intra and interlayer distances obtained from XRD measurements for these crystals: PEA ($d_{intra}$ = 7.3 Å; $d_{inter}$ =12.3Å) > EDA ($d_{intra}$ = 6.7 Å; $d_{inter}$ =9.3Å) > EA ($d_{intra}$ = 6.1 Å; $d_{inter}$ =8.5Å). These results emphasize that the $Co^{2+}$ spin magnetic interaction is not through a direct exchange mechanism, and as the corresponding decrease of the $J/k_B$ value shows, the magnetic coupling becomes weaker. This behavior indicates that individual $Co^{2+}$ spins, slightly oriented, align more easily with the field when they are far away from each other. Therefore, the tetrahedrally coordinated $Co^{2+}$ compounds behave as individual magnetic moments, and their magnetization is favored in the RP-like structures with large Co⋯Co distances.

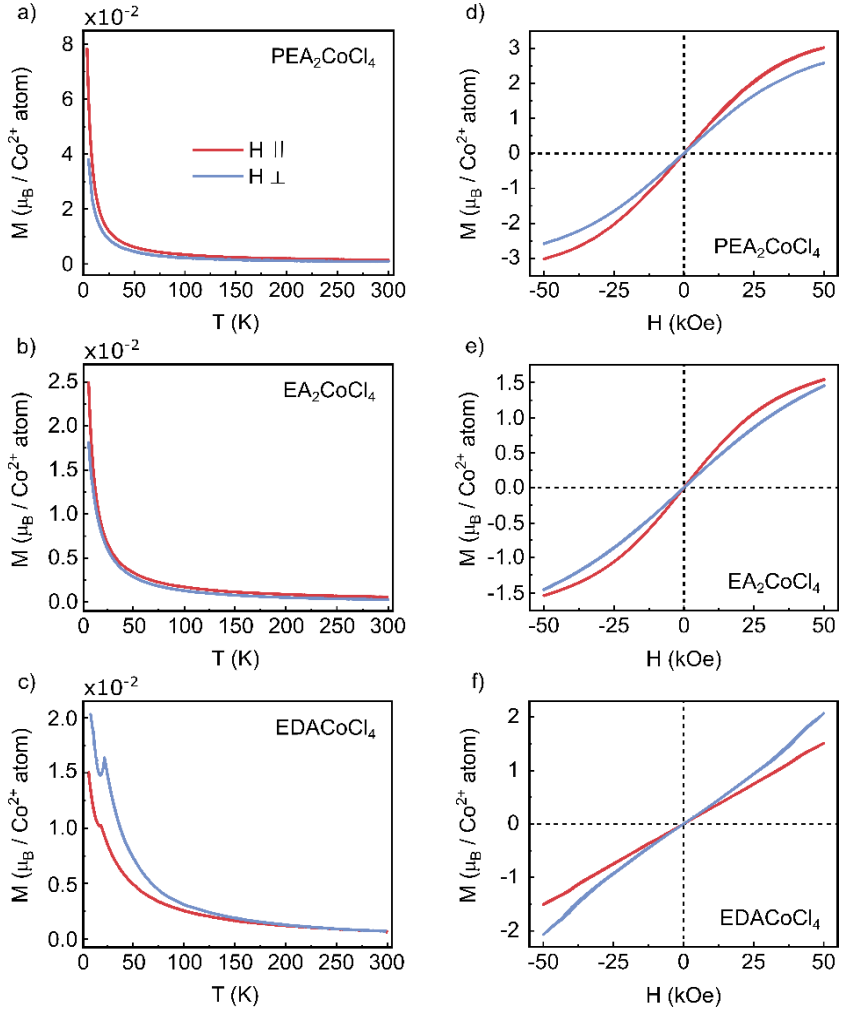

**Figure 5.** Temperature (T) dependent magnetization (M) at 500 Oe parallel (∥, red lines) and perpendicular (⊥, blue lines) to the $[CoCl_4]^{2-}$ tetrahedra layers for (a) $PEA_2CoCl_4$, (b) $EA_2CoCl_4$ and (c) $EDACoCl_4$. Magnetic field (H) dependent magnetization (M) at 5 K parallel and perpendicular to the $[CoCl_4]^{2-}$ tetrahedra layers for (d) $PEA_2CoCl_4$, (e) $EA_2CoCl_4$ and (f) $EDACoCl_4$. Results shown correspond to M(T) and M(H) curves collected from a representative crystal for each material.

To summarize all the above findings, **Table 1** reports the evolution of the magnetic behavior with the applied magnetic field for each layered crystal under study. The table shows how differently the materials behave depending on the transition metal and on the modification of the $M^{2+}$-$M^{2+}$ coupling due to the perovskite phase and the organic cation.

**Table 1.** Summary of the magnetic behavior of the crystals under study classified by transition metal, perovskite phase and organic cation.

| Perovskite phase | Organic cation | Transition metal | | |
|---|---|---|---|---|
| | | $Cu^{2+}$ | $Mn^{2+}$ | $Co^{2+}$ |
| | | Increasing H | Increasing H | Any H |
| RP | PEA$^+$ | FM → FM | AFM with spin-canting → AFM with FM contributions from spin flop transition | paramagnetic |
| RP | EA$^+$ | FM with AFM contributions → FM | AFM → AFM with spin-canting → AFM with FM contributions from spin flop transition | paramagnetic |
| DJ | PEAA$^+$ | AFM with FM contributions → FM | - | - |
| DJ | EDA$^+$ | AFM with FM contributions → AFM | AFM → AFM | paramagnetic |

*RP = Ruddlesden-Popper; DJ = Dion-Jacobson; PEA$^+$ = phenethylammonium; EA$^+$ = ethylammonium; PEAA$^+$ = p-phenylenediammonium; EDA$^+$ = ethylenediammonium; H = magnetic field; FM = ferromagnetic; AFM = antiferromagnetic.

## 3. Conclusions

To sum up, we carried out a detailed study on ten hybrid metal-halide crystals grown by solution-based methods using a series of transition metals ($Cu^{2+}$, $Mn^{2+}$, $Co^{2+}$), while varying both the organic cation nature (alkyl and aryl) and the perovskite phase (RP and DJ). We confirmed their successful synthesis and layered nature by XRD analysis. The study of the vibrational dynamics of these compounds by temperature-dependent Raman spectroscopy revealed that only the samples containing EA$^+$ as organic cation show structural phase transitions below 300K. Moreover, while in the EA$_2$CuCl$_4$ and EA$_2$CoCl$_4$ crystals the structural phase transitions involve the rearrangement of both the inorganic lattice and organic moieties, in the EA$_2$MnCl$_4$ these transitions are governed only by changes in the conformation of the organic moieties. Therefore, in addition to the order-disorder motional freedom of the organic cations, another key factor controlling the structural phase transitions in layered HOIMs is the flexibility of the inorganic metal-halide lattice. Regarding the temperature-dependent photoluminescence on the $Mn^{2+}$ HOIPs, we found that their characteristic photoluminescence emission ~ 600 nm is not significantly affected by the organic cation and perovskite phase, showing that these factors almost do not alter the band structure. Evaluating their magnetic properties, we observed that in the $Cu^{2+}$ HOIPs, a high in-plane anisotropy translated into a deviation of the Cu···X···Cu angles from 180 deg and to a short interlayer distance. This in turn leads to strong AFM contributions from the interlayer interaction and changes from a 2D ferromagnet (PEA$_2$CuCl$_4$) to a 3D antiferromagnet (EDACuCl$_4$). In contrast, the magnetic behavior of the $Mn^{2+}$ HOIPs is intrinsically characterized by AFM intralayer interactions.

However, the appearance and magnitude of the spin-canting, spin-flop transitions and metamagnetic phenomena are also controlled by the anisotropy generated in the crystal structure. A particular case is that of EA$_2$MnCl$_4$, which shows all the mentioned magnetic phenomena. Finally, Co$^{2+}$ HOIMs constituted by unconnected tetrahedra show a paramagnetic behavior, favored in the RP-like perovskite structures due to the large Co···Co distances.

To conclude, our comprehensive study demonstrates that the composition and structural flexibility of hybrid organic-inorganic metal-halides can offer valuable and interesting magnetic properties, opening new pathways towards the design and pursuit of novel layered magnetic materials for optoelectronic and spintronic applications.

## 4. Experimental Section

*Synthesis of the layered HOIMs.* The different crystals used in this work were synthesized solubilizing the corresponding precursors at high temperature and then cooling down to promote the precipitation but in different solvent media depending on the organic cation and perovskite phase, Ruddlesden-Popper (RP) or Dion-Jacobson (DJ).[18]

*Cu-based HOIPs:* The **RP (C$_6$H$_5$CH$_2$CH$_2$NH$_3$)$_2$CuCl$_4$ crystals** were synthesized using a mixture of polar solvents in the medium. Briefly, we mixed CuCl$_2$ (67.2 mg, 99.999% trace metal basis, anhydrous, Sigma Aldrich), and phenylethylamine (PEA-126 µL, ≥99% Sigma Aldrich) in 650 µL of HCl (37% vol, Sigma Aldrich). Then, 2 mL of acetone (≥99.5%, Sigma Aldrich) and 1 mL of ethanol (absolute, suitable for HPLC, ≥99.8%, Sigma Aldrich) were added. We passed a N$_2$ flow during 2 min. Subsequently, the solution was stirred at 150°C up to dissolution. Then, the vial was quickly transferred to another hot plate set at 35°C to allow the growth of the crystals. After 1 day growth, we obtained mm-size yellow-brownish crystals. The **RP (C$_2$H$_5$NH$_3$)$_2$CuCl$_4$ crystals** were grown in a similar way but using as precursor the organic salt instead of the amine, a more polar medium[80] and without adding HCl. We mixed CuCl$_2$ (67.2 mg) and ethylammonium chloride (EACl-81.5 mg, 98% Sigma Aldrich) in 1 mL of methanol (≥99.9%, HPLC quality, Sigma Aldrich) and 1 mL of ethanol. We passed a N$_2$ flow during 2 min. Subsequently, the solution was stirred at 120°C up to dissolution. Then, the vial was quickly transferred to another hot plate set at 35°C to allow the growth of the crystals for 2 days. The **DJ (NH$_3$C$_2$H$_2$NH$_3$)CuCl$_4$ and (NH$_3$C$_6$H$_4$NH$_3$)CuCl$_4$ crystals** were using acid medium[81,82] and setting a temperature ramp 220ºC → 90ºC → 50ºC (rate 20ºC/h) due to the insolubility of the organic cations in polar solvents. We mixed CuCl$_2$ (67.2 mg), and ethylenediamine (EDA-33.4 µL, ≥99% Sigma Aldrich) or p-phenylenediamine (PEAA-54.1 mg, ≥99% Sigma Aldrich) in 4 mL of HCl. The solution was stirred at 240°C up to dissolution. Then, the temperature was progressively decreased up to 50°C to allow the growth of the crystals for 1 day.

*Mn-based HOIPs:* The **($C_6H_5CH_2CH_2NH_3$)$_2$MnCl$_4$ crystals** were synthesized in a similar way than the Cu counterpart using polar solvents in the medium used for growing this material as reported in literature[37]. Briefly, we mixed MnCl$_2$ (63 mg, 99.99%, anhydrous beads 10 mesh, Sigma Aldrich), and PEA (126 µL) in 650 µL of HCl. Then, 2 mL of acetone and 1 mL of ethanol were added, and the solution was stirred at 120°C up to achieve a clear almost transparent solution. Then, the vial was immediately transferred to another hot plate set at 35°C to allow the growth of the crystals. After 1h the first pale pink crystals were formed, but we left growing 1 day. The **RP ($C_2H_5NH_3$)$_2$MnCl$_4$ crystals** were grown in a similar way but using as precursor the organic salt instead of the amine and without adding HCl. We mixed MnCl$_2$ (63 mg) and EACl (81.5 mg) in 1 mL of ethanol. Then, the solution was stirred at 120°C up to dissolution and quickly transferred to another hot plate set at 35°C to allow the growth of the crystals for 2 days. Similarly to the Cu counterpart, the **DJ ($NH_3C_2H_2NH_3$)MnCl$_4$ crystals** were using acid medium[81,82] and setting a temperature ramp 220ºC → 90ºC → 50ºC. We mixed MnCl$_2$ (63 mg), and EDA (33.4 µL) in 4 mL of HCl. The solution was stirred at 240°C up to dissolution. Then, the temperature was progressively decreased to 50°C to allow the growth of the crystals for 1 day. However, we were not successful in obtaining the ($NH_3C_6H_4NH_3$)MnCl$_4$ crystals, probably due to the less flexible Mn-crystal structure compared to Cu.

*Co-based HOIMs:* The **($C_6H_5CH_2CH_2NH_3$)$_2$CoCl$_4$ crystals** were synthesized in a similar way to the Cu and Mn counterparts, but changing the mixture of solvents. Briefly, we mixed CoCl$_2$ (64.9 mg, 99.995% trace metal basis, anhydrous beads, ~10 mesh, Sigma Aldrich), and PEA (126 µL) in 82 µL of HCl in 1.5 mL of methanol (≥99.9%, HPLC quality, Sigma Aldrich) and 1.5 mL of ethanol were added. We passed a N$_2$ flow during 2 min. Subsequently, the solution was stirred at 120°C up to dissolution. Then, the vial was quickly transferred to another hot plate set at 35°C to allow the growth of the crystals. After 7 days growth covering the vial with holey-parchment paper, we obtained blue crystals. The **RP-like ($C_2H_5NH_3$)$_2$CoCl$_4$ crystals** were grown in a similar way as the Cu and Mn counterparts but adding HCl to favor the dissolution of the precursors. We mixed CoCl$_2$ (64.9 mg), EACl (81.5 mg) and 650 µL HCl in 1 mL of methanol and 1 mL of ethanol. Then, the solution was stirred at 120°C up to dissolution and quickly transferred to another hot plate set at 35°C to allow the growth of the crystals for 2 weeks leaving the vial covered with holey-parchment paper. As for the Cu and Mn counterpart, the **DJ-like (NH$_3$C$_2$H$_2$NH$_3$)CoCl$_4$ crystals** were using acid medium[81,82] and setting a temperature ramp 220ºC → 90ºC → 50ºC. We mixed CoCl$_2$ (64.9 mg), and EDA (33.4 µL) in 4 mL of HCl. The solution was stirred at 240°C up to dissolution. Then, the temperature was progressively decreased up to 50°C to allow the growth of the crystals for 7 days. However, as happened for the Mn, we were not successful in obtaining the (NH$_3$C$_6$H$_4$NH$_3$)CoCl$_4$ crystals. In all cases, the grown crystals were isolated by filtration on a Büchner funnel using paper filter (Fisherbrand™ Grade 111 Cellulose) by vacuum suction. Additionally, we dried the collected crystals inside a vial during at least 8h at room temperature in a vacuum Schlenk line. We stored all the crystals in a N$_2$-filled dry box.

*X-ray diffraction characterization:* X-ray diffraction (XRD) patterns were recorded on a Malvern-PANalytical 3$^{rd}$ generation Empyrean X-ray diffractometer, equipped with a 1.8 kW CuKα ceramic X-ray tube, PIXcel3D detector and operating at 45 kV and 40 mA in 0D mode. The diffraction patterns were collected in air at room temperature using Parallel Beam (PB) geometry. XRD data analysis was carried out using HighScore 5.1 software from PANalytical. We confirmed the reproducibility of our results by measuring at least 3 crystals of each material as well as crystals from different batches.

*Raman and Photoluminescence spectroscopy characterization:* Temperature-dependent micro-Raman and photoluminescence (PL) characterization were carried out in a Renishaw® inVia Qontor micro-Raman instrument equipped with a 50× objective (WD 11 mm) and connected to a Linkam® liquid N$_2$ vacuum chamber (10$^{-6}$ hPa). Spectra were collected in a temperature range of 80-340 K using two excitation wavelengths, 532 and 633 nm, with an incident power <1mW both selected to avoid damage to the crystals during the acquisitions. In detail, for PL measurements of Mn$^{2+}$ HOIPs, the 532 nm laser was used. The scattered signal was dispersed by a diffraction grating of 1800 l/mm. Matlab® programming language was employed to analyze the measurements, fitting Raman and PL spectra datasets with Lorentzian and Gaussian functions, respectively. We checked the reproducibility of our results by measuring at least 3 crystals of each material as well as crystals from different batches. The data analysis performed using Matlab® and displayed in the manuscript and SI figures corresponds to the average and standard deviation (error) from 5 different points taken in a representative sample for each material under study. The error found for these measurements is < 1%, with maximum values of ±0.8 cm$^{-1}$ and ±2 cm$^{-1}$ for Raman shift and linewidth, respectively, and ±0.3 nm for PL peak position and FWHM.

*Magnetic characterization*: Magnetic measurements along in-plane and out-of-plane crystal orientations were performed on a Physical Properties Measurement System (PPMS) using the vibrating sample magnetometer mode. Temperature-dependent magnetization was measured between 5 and 300 K at different constant fields as indicated in the corresponding M(T) graphs. Isothermal magnetization at 5 K was carried out with a field sweep range of ±50 kOe. Angle-dependent measurements were carried out cutting the sample into a square-shape and aligning it in the sample holder. We confirmed the reproducibility of our results by measuring at least 3 crystals of each material as well as crystals from different batches. For the transition temperature determination by means of the 1$^{st}$ derivative of the M(T) curves, we found an error of ±2K.

**Supporting Information**
Room-temperature XRD $Mn^{2+}$ and $Co^{2+}$ crystals characterization; additional temperature-dependent micro-Raman and photoluminescence spectroscopy measurements and data analysis; representative fittings examples; and angle-variation and field-dependent magnetic measurements. Supporting Information is available from the Wiley Online Library or from the author.

**Acknowledgements**
This work is supported by the Spanish MCIN/AEI under Projects PID2019-108153GA-I00, RTI2018-094861-B-I00, PID2021-128004NB-C21 and under the María de Maeztu Units of Excellence Programme (Grant CEX2020-001038-M & CEX2020-001067-M). This work is also supported by the FLAG-ERA grant MULTISPIN, by the Spanish MCIN/AEI with grant number PCI2021-122038-2A. This project has received funding from the European Research Council (ERC) under the European Union's Horizon 2020 research and innovation programme (Grant Agreement No. 722951). Additionally, this work was carried out with support from the Basque Science Foundation for Science (IKERBASQUE), POLYMAT, EHU/UPV, Gipuzkoa Council and Basque Government (BERC programme). Technical and human support provided by SGIker of UPV/EHU and European funding (ERDF and ESF) is also acknowledged. B.M-G. thanks Gipuzkoa Council (Spain) in the frame of Gipuzkoa Fellows Program. M.G. acknowledges support from la Caixa Foundation (ID 0010434) for a Junior Leader fellowship (Grant No. LCF/BQ/PI19/11690017). Authors thank R. Llopis (Nanodevices group – CIC nanoGUNE) for his support in the assembly of the Linkam vacuum chamber – Raman instrument set-up; and Dr. E. Goiri Little (Nanodevices group – CIC nanoGUNE) for reading and revising the manuscript.

**Conflict of interest**

The authors declare no conflict of interest.